\def\pr#1 {Phys. Rev. {\bf D#1\tie \rm }}
\def\pe#1 {Phys. Rev. {\bf #1\tie\rm }}
\def\pre#1 {Phys. Rep. {\bf #1\tie\rm }}
\def\pl#1 {Phys. Lett. {\bf #1B\tie \rm }}
\def\prl#1 {Phys. Rev. Lett. {\bf #1\tie \rm }}
\def\np#1 {Nucl. Phys. {\bf B#1\tie \rm }}
\def\ap#1 {Ann. Phys. (NY) {\bf #1\tie \rm }}
\def\cmp#1 {Commun. Math. Phys. {\bf #1\tie \rm }}
\def\imp#1 {Int. Jour. Mod. Phys. {\bf A#1\tie \rm }}
\def\mpl#1 {Mod. Phys. Lett. {\bf A#1\tie\rm }}
\def\jhep#1 {JHEP {\bf #1\tie\rm }}
\def\zp#1 {Z. Phys. {\bf C#1\tie\rm }}

\def\tie{\noexpand~}

\def\be{\begin{equation}}
\def\ee{\end{equation}}
\def\bea{\begin{eqnarray}}
\def\eea{\end{eqnarray}}

\catcode`@=11
\def\marginnote#1{}
\newcount\hour
\newcount\minute
\newtoks\amorpm
\hour=\time\divide\hour by60
\minute=\time{\multiply\hour by60 \global\advance\minute by-\hour}
\edef\standardtime{{\ifnum\hour<12 \global\amorpm={am}%
        \else\global\amorpm={pm}\advance\hour by-12 \fi
        \ifnum\hour=0 \hour=12 \fi
        \number\hour:\ifnum\minute<10 0\fi\number\minute\the\amorpm}}
\edef\militarytime{\number\hour:\ifnum\minute<10 0\fi\number\minute}
\def\draftlabel#1{{\@bsphack\if@filesw {\let\thepage\relax
   \xdef\@gtempa{\write\@auxout{\string
      \newlabel{#1}{{\@currentlabel}{\thepage}}}}}\@gtempa
   \if@nobreak \ifvmode\nobreak\fi\fi\fi\@esphack}
        \gdef\@eqnlabel{#1}}
\def\@eqnlabel{}
\def\@vacuum{}
\def\draftmarginnote#1{\marginpar{\raggedright\scriptsize\tt#1}}
\def\draft{\oddsidemargin 0.0truein
        \def\@oddfoot{\sl preliminary draft \hfil
        \rm\thepage\hfil\sl\today\quad\militarytime}
        \let\@evenfoot\@oddfoot \overfullrule 3pt
        \let\label=\draftlabel
        \let\marginnote=\draftmarginnote
   \def\@eqnnum{(\theequation)\rlap{\kern\marginparsep\tt\@eqnlabel}%
\global\let\@eqnlabel\@vacuum}  }
\catcode`@=12

\documentstyle[epsf, 12pt]{article}
\begin{document}

\thispagestyle{empty}
\begin{flushright}
RU04-06-B
\end{flushright}

\bigskip\bigskip
\begin{center}
\Large{\bf CHROMOMAGNETIC INSTABILITY AND THE LOFF STATE\\[2mm]
IN A TWO FLAVOR COLOR SUPERCONDUCTOR}
\end{center}

\vskip 1.0truecm

\centerline{\bf Ioannis
Giannakis and Hai-Cang Ren\footnote{giannak@summit.rockefeller.edu,
ren@summit.rockefeller.edu}}
\vskip5mm
\centerline{\it Physics Department}
\centerline{\it Rockefeller University}
\centerline{\it 1230 York Avenue}
\centerline{\it New York, NY 10021}

\vskip5mm

\bigskip \nopagebreak \begin{abstract}
\noindent
We explore the relation between the chromomagnetic instability of a 
homogeneous two flavor color superconductor and the LOFF state.
We perturb the free energy of the 2SC by generating
a small net momentum for the quark pair. 
We find that the imaginary
Meissner mass of a particular gluon implies that the LOFF
state is energetically favored.

\end{abstract}

\newpage\setcounter{page}1

\vfill\vfill\break

It is well-known that nuclear matter when is subjected to
extreme conditions of temperature and density undergoes phase
transitions. New forms of matter have long been sought after,
both theoretically and experimentally. Under sufficiently high temperature,
nuclear matter is expected to undergo a phase transition to a quark-gluon
plasma. At low temperatures but sufficiently high density, however, 
recent studies suggest a much
richer phase structure. One of these phases-color superconductivity-
might exist in the core of neutron stars \cite{color, reviews}.
The baryon density of a color superconductor appropriate to a 
neutron star is few times higher than the saturation density
of nuclear matter. Consequently, perturbative QCD becomes
inadequate since the Fermi energy of the quarks is not much
larger than $\Lambda_{\rm QCD}$.

An additional complication emerges at the baryon density appropriate for 
a compact star. The mass of the strange quark $s$ can no longer be ignored.
This causes
mismatches between the Fermi sea of the different quark 
flavors. A large number of
exotic Color Superconducting (CSC) phases has been investigated in
the literature: these include the CFL ( color 
flavor locking ) state \cite{alford}, the LOFF state \cite{loff},
the 2SC ( two flavor pairing ) and g2SC
( gapless 2SC ) states \cite{shovkovy} \cite{chao}, the
gCFL ( gapless CFL ) state \cite{kouvaris} and super-normal 
mixed states \cite{bedaque}. 
The stability of these phases becomes an important issue to address. 
The interplay of various homogeneous phases has been investigated recently
\cite{ruester}.

Gapless color superconductivity, g2SC or gCFL, is a consequence of breached 
pairing among the different species of fermions. It was considered 
initially in the context of an electronic 
superconductor by Sarma \cite{sarma}, and recently 
it was applied by Liu and Wilczek
in the context of cold atomic gas  \cite{liu}.
If the force responsible for pairing exceeds a
critical value, the gap equation for fixed
chemical potential for each species admits two branches 
of solutions. One with fully gaped excitation spectrum 
and another gapless one.
But if we impose the constraint that the total number of 
each species is fixed, or other constraints such 
as the charge neutrality \cite{shovkovy} etc.,
the gapless superconductor becomes the only solution. 
One problem that arises with the gapless solution is the possibility of 
imaginary Meissner masses \cite{wu}.
Recently, Huang and Shovkovy found that the squares of the chromomagnetic 
masses of the five gluons corresponding to the broken
generators of the $SU(3)_{c}$ group are negative at
zero temperature \cite{huang}. This result indicates the existence
of a chromomagnetic type plasma instability for the
underlying color superconducting phase.
At zero temperature, this instability extends over the entire branch
of the gapless solution and persists in the 2SC phase whenever the mismatch
between the Fermi momenta of the different quark flavors is larger
than a critical value. Such a chromomagnetic instability has also been 
discovered in the gCFL phase \cite{casal}. The relation of the chromomagnetic 
instability and the mixed state were explored in Ref. \cite{rebby}.

In this note, we shall examine the relationship between 
the chromomagnetic instability
and the LOFF state. For a 2SC in particular, we find that
the chromomagnetic instability associated with the eighth gluon is a
sufficient condition for a LOFF state of lower free energy.
Complementing
the results of \cite{huang}, we calculate the Meissner screening masses
in the region of the phase diagram of QCD, near
the second order phase transition to the normal phase. We find that the 
value of the mismatch at the onset of the chromomagnetic instability 
agrees with the threshold value of the LOFF state
along the transition line, know to an electronic 
superconductor \cite{tanaka}.

We consider the pairing of two quark flavors, $u$ and $d$.
The NJL effective Lagrangian of two flavors coupled to
external color and 
electromagnetic gauge fields in the massless limit is given
by
\begin{eqnarray}
&{\cal L}& = -\bar\psi\gamma_\mu\Big(\frac{\partial}{\partial x_\mu}
-igA_\mu^lT_l-ieQA_\mu\Big)\psi+\bar\psi\gamma_4\mu\psi\nonumber \\
&+&G_S[(\bar\psi\psi)^2+(\bar\psi\vec\tau\psi)^2]
+G_D(\bar\psi_C\gamma_5\epsilon^c\tau_2\psi)
(\bar\psi\gamma_5\epsilon^c\tau_2\psi_C)
\label{NJL},
\end{eqnarray}
where $\psi$ is the quark field, $A_\mu^l$ the gluon potential
and $A_\mu$ the
electromagnetic potential. The four fermion interaction terms of 
(\ref{NJL}) is taken from Ref. \cite{hzc}, where $(\epsilon^c)^{mn}=\epsilon^{cmn}$ is 
a $3\times 3$ matrix acting on the red, green and blue color indices. 
The lagrangian is invariant under $SU(3)_c\times U(1)_{em}$ 
transformations. The $SU(3)_c$ generators and the electromagnetic
charge operator are written as
\begin{equation}
T_l=\frac{1}{2}\lambda_l, \qquad Q=\frac{1}{6}+\frac{1}{2}\tau_3
\label{charge}
\end{equation}
where $\lambda_l$ and $\vec\tau$ are
the Gell-Mann and Pauli matrices respectively.
The chemical potential reads $\mu=\bar\mu
-\delta\tau_3+\delta^\prime\lambda_8$,
with $\delta$ and $\delta^\prime$ being sensitive to the charge
and color neutrality respectively. All gamma matrices are hermitian. 

The CSC order parameter is set to be  
\begin{equation}
<\bar\psi_f^c\gamma_5(\psi_C)_{f^\prime}^{c^\prime}>=\Phi_{ff^\prime}^{cc^\prime},
\label{order}
\end{equation}
where $\Phi=-\frac{\Delta}{8G_D}
\lambda_2\tau_2$ is a matrix both in color and flavor space, i.e.
\begin{equation}
\Phi_{f^\prime f}^{c^\prime c}=
i\frac{\Delta}{8G_D}\epsilon^{c^\prime cb}\epsilon_{f^\prime f}
\end{equation}
with $\Delta>0$, $f=u, d$ and $c=r,g,b$. 
By ignoring the chiral condensate and expanding the NJL
Lagrangian to linear order in the fluctuation
\begin{equation}
\bar\psi_f^c\gamma_5(\psi_C)_{f^\prime}^{c^\prime}
-\Phi_{ff^\prime}^{cc^\prime},
\label{order}
\end{equation}
we derive the following mean field expression for the NJL lagrangian
\begin{equation}
{\cal L}_{MF}=-\frac{\Delta^2}{4G_D}
-\bar\psi\gamma_\mu\Big(\frac{\partial}{\partial x_\mu}-igA_\mu^lT^l
-ieQA_\mu\Big)\psi+\bar\psi\gamma_4\mu\psi
+\Delta[-\bar\psi_C\gamma_5\lambda_2\tau_2\psi
+\bar\psi\gamma_5\lambda_2\tau_2\psi_C].
\label{NJL_MF}
\end{equation}
The thermodynamic potential $\Omega$ is given by 
the Euclidean path integral 
\begin{equation}
\Omega=k_BT\ln\Big[\int[d\psi d\bar\psi]
\exp\Big(\int d^4x{\cal L}_{MF}\Big)\Big]
\end{equation}
with $0<x_4<(k_BT)^{-1}$ and can be expanded in terms of one loop diagrams.
The corresponding free energy, ${\cal F}\equiv-\Omega$ will be minimized 
at equilibrium. 

The Nambu-Gorkov form of the
inverse quark propagator in the super phase takes the form
\begin{equation}
{\cal S}^{-1}(P)=\left( \begin{array}{cc} 
/\kern-7pt P+\mu\gamma_4 & \Delta\gamma_5\lambda_2\tau_2 \\ 
-\Delta\gamma_5\lambda_2\tau_2 & /\kern-7pt P-\mu\gamma_4
\end{array} \right)
\label{prop_gap}
\end{equation}
where $/\kern-7pt P=-i\gamma_\mu P_\mu$, $P=(\vec p,-\nu)$ and
$\nu=(2n+1)\pi k_BT$ is the Matsubara frequency. $\psi_C=C\tilde{\bar\psi}$ 
is the charge conjugate of $\psi$ with $C=i\gamma_2\gamma_4$. 
Each block of the matrix quark propagator (\ref{prop_gap}) is 
itself a matrix in
color-flavor space.
The NG representation of the quark-gluon and quark-photon vertex is 
\begin{equation}
\Gamma_\mu^l =\left( \begin{array}{cc}
\gamma_\mu {\cal T}_l & 0 \\
 0 & -\gamma_\mu\tilde {\cal T}_l
\end{array} \right),
\end{equation}
with $l=1,2,...,9$ and the tilde standing for transpose. We have
\begin{equation}
{\cal T}_l=T_l
\end{equation}
for $l=1,...,7$,
\begin{equation}
{\cal T}_8=T_8\cos\theta+\frac{e}{g}Q\sin\theta
\end{equation}
and
\begin{equation}
{\cal T}_9=-\frac{g}{e}T_8\sin\theta+Q\cos\theta
\end{equation}
with 
$\cos\theta=\frac{\sqrt{3}g}{\sqrt{3g^2+e^2}}$ and $\sin\theta
=\frac{e}{\sqrt{3g^2+e^2}}$. 
Here we have introduced the gluon-photon mixing \cite{mixing}. 
The condensate is invariant under the residual gauge transformations 
$SU(2)\times{\cal U}(1)$, where $SU(2)$ is generated by  
$T_1$, $T_2$, $T_3$ and ${\cal U}(1)$
by ${\cal T}_9$. 

A particularly convenient representation of the quark 
propagator and the vertex can be obtained by performing
a unitary transformation generated by
\begin{equation}
U=\left( \begin{array}{cc} 1 & 0\\ 
0 & \tau_2\end{array} \right).
\label{unitary}
\end{equation}
We have
\begin{eqnarray}
{\cal S}^{\prime -1}(P) &\equiv& U{\cal S}^{-1}(P)U^\dagger\nonumber \\
&=& \left( \begin{array}{cc} 
/\kern-7pt P+\bar\mu\gamma_4-\delta\gamma_4\tau_3+\delta^\prime\lambda_8 & \Delta\gamma_5\lambda_2 \\ 
-\Delta\gamma_5\lambda_2 & /\kern-7pt P-\bar\mu\gamma_4-\delta\gamma_4\tau_3-\delta^\prime\lambda_8
\end{array} \right)
\label{prop_gap_new}
\end{eqnarray}
and 
\begin{equation}
\Gamma_\mu^{\prime l} =\left( \begin{array}{cc}
\gamma_\mu T_l & 0 \\
 0 & -\gamma_\mu\tau_2\tilde T_l\tau_2
\end{array} \right).
\end{equation}
In the remaining of this paper we shall use this representation
and we shall suppress the primes from
${\cal S}$ and $\Gamma$. We notice that ${\cal S}(P)$ in 
the new representation commutes with $\tau_3$.

The squared Meissner mass is defined by 
\begin{equation}
(m^2)^{l^\prime l}=\frac{1}{2}\lim_{\vec K\to 0}(\delta_{ij}-\hat k_i\hat k_j)
\Pi_{ij}^{l^\prime l}(K)
=\frac{1}{2}(\delta_{ij}-\hat z_i\hat z_j)\Pi_{ij}^{l^\prime l}(0)
\end{equation}
where 
\begin{equation}
\Pi_{ij}^{l^\prime l}(K)=-\frac{k_BT}{2}\int\frac{d^3\vec p}{(2\pi)^3}
Tr\Gamma_i^{l^\prime}{\cal S}(P)\Gamma_j^{l}{\cal S}(P+K)
\label{eqkit}
\end{equation}
is the gluon polarisation tensor with
$K=(\vec k,-\omega)$. Furthermore we
have $(m^2)^{l^\prime l}=0$ for $\Delta=0$.

A simplification follows from the observation that the isospin part of the 
charge operator (\ref{charge}) contributes to the Meissner masses
a total derivative,
i.e.
\begin{equation}
-\frac{k_BT}{2}\int\frac{d^3\vec p}{(2\pi)^3}
Tr\gamma_j\tau_3{\cal S}(P)M{\cal S}(P)
= -i\frac{k_BT}{2}\int\frac{d^3\vec p}{(2\pi)^3}
\frac{\partial}{\partial p_j}Tr\tau_3M{\cal S}(P)
\end{equation}
where $M$ is an arbitrary matrix. Its contribution
then can be omitted and the charge operator may be effectively replaced
by $Q=\frac{1}{6}$.

It follows from the residual $SU(2)$ symmetry that
$(m^2)^{l^\prime l}$ is diagonal
with respect to $\l^\prime$ and $l$ that label the standard 
Gell-Mann matrices.
We proceed to divide the set of the nine generators into three 
subsets.
Subset I includes the unbroken generators ${\cal T}_1$, 
${\cal T}_2$ and ${\cal T}_3$,
subset II includes the broken generators ${\cal T}_4$, 
${\cal T}_5$, ${\cal T}_6$
and ${\cal T}_7$ while subset III includes the broken ${\cal T}_8$ and the 
unbroken ${\cal T}_9$ generators. Under an $SU(2)$ transformation,
${\cal T}_l\to u{\cal T}_lu^\dagger$,
the generators in the subset I transform as the adjoint representation, 
the generators in the subset II as the fundamental 
one while the generators in the subset III 
remain invariant. Therefore the indices $l^\prime$ 
and $l$ which correspond to generators from different subsets
do not mix in the expression for $(m^2)^{l^\prime l}$. The mass 
matrix within subset I
vanishes following the standard arguments of gauge invariance.
As to the subset II, we proceed to relabel its members according 
to ${\cal J}_1=T_4$, ${\cal J}_{\bar 1}
=T_5$, ${\cal J}_2=T_6$ and ${\cal J}_{\bar 2}=T_7$. Introducing 
\begin{equation}
\phi_1= \left( \begin{array}{cc} 1 \\ 0 \end{array} \right) \hbox{\\\\\\\\}
\qquad \phi_2= \left( \begin{array}{cc} 0 \\ 1\end{array} \right),
\end{equation}
we have 
\begin{equation}
{\cal J}_\alpha=\left( \begin{array}{cc} 0 
& \phi_\alpha \\ \phi_\alpha^\dagger & 0 \end{array} \right),
\end{equation}
and similar expressions for ${\cal J}_{\bar\alpha}$ with 
$\bar\phi_\alpha=-i\phi_\alpha$, where 
the block decomposition is with respect to color indices.
Since ${\cal J}_\alpha$ is symmetric and
${\cal J}_{\bar\alpha}$ is antisymmetric, there is
no mixing between ${\cal J}_\alpha$ and
${\cal J}_\beta$, following from the properties 
of the matrix $C=i\gamma_2\gamma_4$, i.e.
$C\tilde\gamma_\mu C^{-1}=-\gamma_\mu$ and $C\tilde{\cal S}(P)C^{-1}
=-{\cal S}(P)$. The
matrix element $(m^{2})^{\alpha\beta}$ or $(m^{2})^{\bar\alpha\bar\beta}$ 
can be written as a sum of terms $(\phi^\alpha)^\dagger M\phi^\beta$ or
$(\phi^{\bar\alpha})^\dagger M\phi^{\bar\beta}$ with
the same $SU(2)$ invariant $M$. Therefore we have $(m^{2})^{\alpha\beta}
=(m^{2})^{\bar\alpha\bar\beta}\propto\delta_{\alpha\beta}$.  
Finally in subset III, the color matrices
never mix the red and green components with the 
blue ones. The latter do not carry a condensate. It follows 
then that we need only to work 
within the red-green subspace, in which ${\cal T}^9$ does not 
contribute and ${\cal T}^8$ 
may be replaced by a factor $\frac{1}{6}\sqrt{3g^2+e^2}$.   

Next we proceed to define the susceptibility as the response of the 
free energy to a small net  momentum of the Cooper pair.
This amounts in replacing the order parameter (\ref{order}) with 
\begin{equation}
<\bar\chi\gamma_5\chi_C>=\Phi
\label{eqorder}
\end{equation}
where
\begin{equation} 
\chi(\vec r)=e^{i\vec q\cdot\vec r}\psi(\vec r)
\label{loff}
\end{equation}
and to form the NG basis with respect to the new field $\chi$ 
and its charge conjugate 
$\chi_C=C\tilde{\bar\chi}$. The net momentum carried by the Cooper 
pair is then $2\vec q$.
The NJL lagrangian (\ref{NJL_MF}) expressed in terms of $\chi$ reads
\begin{eqnarray}
&{\cal L}& = -\bar\chi\gamma_\mu\Big(\frac{\partial}{\partial x_\mu}
-igA_\mu^lT^l-ieQA_\mu\Big)\chi-i\bar\chi\vec\gamma\cdot\vec q\chi
+\bar\chi\gamma_4\mu\chi \nonumber \\
&+& G_S[(\bar\chi\chi)^2+(\bar\chi\vec\tau\chi)^2]
+G_{D}(\bar\chi_C\gamma_5\epsilon^c\tau_2\chi)
(\bar\chi\gamma_5\epsilon^c\tau_2\chi_C)
\label{NJL},
\end{eqnarray}
and gives rise to 
a new vertex $\vec q\cdot\vec\Gamma$ with
\begin{equation}
\vec\Gamma =\left( \begin{array}{cc}
\vec\gamma & 0 \\
 0 & -\vec\gamma
\end{array} \right)
\end{equation}
in NG basis. The free energy can be expanded according 
to increasing powers of $q$,
\begin{equation}
{\cal F}={\cal F}_0+\frac{1}{2}\kappa q^2+....
\label{pert}
\end{equation}

We call the coefficient $\kappa$ momentum susceptibility. 
A signal that the LOFF state has lower free energy than the 2SC state
is a negative 
momentum susceptibility.
We have
\begin{equation}
\kappa = -\frac{1}{3}\delta_{ij}\frac{k_BT}{2}\int\frac{d^3\vec p}{(2\pi)^3}
Tr\Gamma_i{\cal S}(P)\Gamma_j{\cal S}(P).
\end{equation}
We observe that the relationship
\begin{equation}
m_8^2=\frac{1}{12}\Big(g^2+\frac{e^2}{3}\Big)\kappa
\end{equation}
is valid for all temperatures in the super phase. 
Therefore the chromomagnetic instability of the gluon that corresponds to the
8th color generator is equivalent to the instability against LOFF
pairing.

In order to compare our results with the existing literature
on LOFF pairing \cite{tanaka},
we calculate explicitly the
Meissner masses and the momentum susceptibility near the second order 
transition temperature where $\Delta<<k_BT$. We shall make 
the approximation that 
$\delta^\prime = 0$, consistent with the numerical study in Ref.
\cite{shovkovy}. As a result $\bar\mu=\frac{1}{2}(\mu_u+\mu_d)$ and 
$\delta=\frac{1}{2}(\mu_d-\mu_u)$.
While it is straightforward to obtain a closed form of 
the quark propagator for an 
arbitrary gap, it will suffice to expand it to second 
order in $\Delta$, with 
$\Delta$ being the solution of the gap equation as $T\to T_c$. After 
some algebra, we find that
\begin{equation}
(m^2)^{l^\prime l}=\frac{1}{2}g^2\Delta^2[C_1^{l^\prime l}I
+C_2^{l^{\prime}l}J].
\label{eqwar}
\end{equation}
The quantities $I$ and $J$ are given by
\begin{eqnarray}
I = -\frac{1}{2}k_BT\sum_\nu\int\frac{d^3\vec p}{(2\pi)^3}
(\delta_{ij} & -& \hat k_i\hat k_j){\rm tr}
\Big [
\gamma_i S_+^u(P)\gamma_jS_+^u(P)
S_-^d(P)S_+^u(P) \nonumber \\
& + & \gamma_iS_+^d(P)\gamma_jS_+^d(P)S_-^u(P)S_+^d(P)\Big].
\label{eqrat}
\end{eqnarray}
and
\begin{eqnarray}
J =  \frac{1}{4}k_BT\sum_\nu\int\frac{d^3\vec p}{(2\pi)^3}(\delta_{ij}
& - & \hat k_i\hat k_j){\rm tr}
\Big[
\gamma_i S_+^u(P)S_-^d(P)\gamma_j
S_-^d(P)S_+^u(P) \nonumber \\
& + & \gamma_iS_+^d(P)S_-^u(P)\gamma_jS_-^u(P)
S_+^d(P)\Big],
\label{eqcion}
\end{eqnarray}
with
\begin{equation}
S_\pm^f(P)=\frac{1}{/\kern-7pt P\pm\mu_f\gamma_4}
\end{equation}
and $S_\mp^f(P)=-S_\pm^f(-P)$. Using the relation
\begin{equation}
\frac{\partial}{\partial p_j}S_\pm^f(P)=iS_\pm^f(P)\gamma_jS_\pm^f(P)
\label{deriv}
\end{equation}
and integrating by parts, we can show that $I=J$. 

The group theoretic factors
\begin{equation}
C_1^{l^\prime l}=4{\rm tr}({\cal T}_{l^\prime}{\cal T}_l
+{\cal T}_l{\cal T}_{l^\prime})\lambda_2^2
\end{equation}
and
\begin{equation}
C_2^{l^\prime l}=4{\rm tr}{\cal T}_{l^\prime}\lambda_2
\tilde {\cal T}_l\lambda_2
\end{equation}
in the standard representation of the Gell-Mann 
matrices are given by
\begin{equation}
C_1={\rm diag.}\Big(2,2,2,1,1,1,1,\frac{2}{3}\Big(1+\frac{e^2}{3g^2}\Big),0\Big)
\label{eqbios}
\end{equation}
and
\begin{equation}
C_2={\rm diag.}\Big(-2,-2,-2,0,0,0,0,\frac{2}{3}\Big(1+\frac{e^2}{3g^2}\Big),0\Big).
\label{eqzoi}
\end{equation}
Combining Eq. (\ref{eqwar}),  (\ref{eqbios}) and
(\ref{eqzoi}), we obtain the following expressions for the Meissner
masses of the gluons
\begin{equation}
m^2\vert_{4,5,6,7}=\frac{1}{2}g^2\Delta^2J,
\end{equation}
\begin{equation}
m^2\vert_8=\frac{2}{3}\Big(g^2+\frac{e^2}{3g^2}\Big)\Delta^2J.
\end{equation}
correponding to the 
broken generators ${\cal T}_4$, ${\cal T}_5$, ${\cal T}_6$, ${\cal T}_7$ 
and ${\cal T}_8$ of the $SU(3)_{c}\times U(1)_{\rm em}$ respectively.

Carrying out the momentum integral in Eq.(\ref{eqcion}), we obtain the
expression
\begin{eqnarray}
J&=& \frac{\mu^2k_BT}{3\pi}{\rm Re}\Big[\sum_{\nu>0}\frac{1}{(\nu+i\delta)^3}
+\frac{3}{\mu^2}\sum_{0<\nu<\Lambda}
\frac{1}{\nu+i\delta}\Big]\nonumber \\
&=& -\frac{\mu^2}{24\pi^4k_B^2T^2}{\rm Re}\psi^{\prime\prime}\Big(\frac{1}{2}
+i\frac{\delta}{2\pi k_BT}\Big) \\ \nonumber
&-&\frac{1}{2\pi^2}\Big[{\rm Re}\psi\Big(\frac{1}{2}+i\frac{\delta}
{2\pi k_BT}\Big)
+\gamma_E-\sum_{n=1}^N\frac{1}{n} \Big],
\end{eqnarray}
where $\psi(z)=\frac{d\ln\Gamma(z)}{dz}$, $\gamma_E=0.5772...$ and we have 
introduced an UV cutoff $\Lambda$ in the summation 
over the Matsubara energies with $\Lambda=2\pi k_BT\Big(N+\frac{1}{2}\Big)$. 
$\Lambda$ represents the energy scale of the Cooper pair. Until this point
we have made no assumptions
about the size of the mismatch $\delta$ when compared to the 
strength of the average chemical potential $\bar\mu$.
The typical value of $\delta$ that results from the charge 
neutrality condition is not much smaller
than $\bar\mu$ \cite{shovkovy}.  

For $\delta<<\mu$, $\Lambda<<\mu$ and $k_BT<<\mu$, we have 
\begin{equation}
J \simeq -\frac{\mu^2}{48\pi^4k_B^2T^2}{\rm Re}\psi^{\prime\prime}
\Big(\frac{1}{2}+i\frac{\delta}{2\pi k_BT}\Big)
\end{equation}
and the term dropped is smaller by an order of $O\Big(\Big(\frac
{k_BT}{\mu}\Big)^2\ln\frac{\mu}{\Lambda}
\Big)$. The Meissner masses of the gluons correponding to the 
broken generators ${\cal T}_4$, ${\cal T}_5$, ${\cal T}_6$ and ${\cal T}_7$ 
are given by
\begin{equation}
m^2\vert_{4,5,6,7}=-\frac{g^2\mu^2\Delta^2}{96\pi^4(k_BT)^2}
{\rm Re}\psi^{\prime\prime}\Big(\frac{1}{2}+i\frac{\delta}{2\pi k_BT}\Big),
\end{equation}
while the Meissner mass for the gluon corresponding
to the broken generator ${\cal T}_8$ by
\begin{equation}
m^2\vert_8=-\frac{g^2\mu^2\Delta^2}
{72\pi^4(k_BT)^2}\Big(1+\frac{e^2}{3g^2}\Big)
{\rm Re}\psi^{\prime\prime}\Big(\frac{1}{2}+i\frac{\delta}{2\pi k_BT}\Big).
\end{equation}
The momentum susceptibility under the same approximation reads
\begin{equation}
\kappa =-\frac{\mu^2\Delta^2}{6\pi^4(k_BT)^2}
{\rm Re}\psi^{\prime\prime}\Big(\frac{1}{2}+i\frac{\delta}{2\pi k_BT}\Big).
\end{equation}
In the absence of the mismatch, $\delta=0$, we recover,
with the aid of the formula
$\psi^{\prime\prime}\Big(\frac{1}{2}\Big)
=-14\zeta(3)$, the familiar parameters 
of the Ginzburg-Landau free energy \cite{iida},\cite{GR},
\begin{equation}
m^2\vert_{4,5,6,7}=\frac{7\zeta(3)g^2\mu^2\Delta^2}{48\pi^4(k_BT)^2},
\end{equation}
\begin{equation}
m^2\vert_8=\frac{7\zeta(3)g^2\mu^2\Delta^2}
{36\pi^4(k_BT)^2}\Big(1+\frac{e^2}{3g^2}\Big)
\end{equation}
and
\begin{equation}
\kappa=\frac{7\zeta(3)\mu^2\Delta^2}{3\pi^4(k_BT)^2}.
\end{equation}
The real part of the function -$\psi^{\prime\prime}\Big(\frac{1}{2}
+i\frac{\delta}{2\pi k_BT})$ is 
positive for small mismatch and changes its sign
at $\frac{\delta}{2\pi k_BT}\simeq 0.3041$.
This is the critical value which signals the emergence of the chromomagnetic
instability. Concurrently, the momentum susceptibility
becomes negative signaling the LOFF instability.

The pairing temperature for a mismatched Fermi sea of electrons at 
given $\delta$ and $q$ has been worked out and is given by the solution to 
the transcendental equation \cite{tanaka}
\begin{equation}
\ln\frac{T_{\delta,q}}{T_0}=\frac{\pi k_B T_{\delta,q}}{q}
{\rm Im}\ln\frac{\Gamma\Big(\frac{1}{2}-i\frac{\delta+q}{2\pi k_BT_{\delta,q}}
\Big)}{\Gamma\Big(\frac{1}{2}-i\frac{\delta-q}{2\pi k_BT_{\delta,q}}
\Big)}-\gamma_E-2\ln 2.
\end{equation}
The transition temperature for a given $\delta$
is the maximum of $T_{\delta, q}$
with respect to $q$. The small $q$ expansion of $T_{\delta,q}$ reads
\begin{eqnarray}
\ln\frac{T_{\delta,q}}{T_c}& = & -\gamma_E-2\ln 2-{\rm Re}\psi\Big(\frac{1}{2}
+i\frac{\delta}{2\pi k_BT_{\delta,0}}\Big)\nonumber \\
& + & \frac{q^2}{24\pi^2 k_B^2T_{\delta, 0}^2}{\rm Re}\psi^{\prime\prime}
\Big(\frac{1}{2}+i\frac{\delta}{2\pi k_BT_{\delta, 0}}\Big)+ \cdots
\end{eqnarray}
The LOFF pairing prevails when the coefficient of $q^2$ becomes positive, 
which is exactly the point where
the expression for the Meissner masses changes 
sign. Upon substitution of the value
$\frac{\delta}{2\pi k_BT_{\delta,0}}\simeq 0.3041$, we obtain that
\begin{equation}
\frac{\delta}{\Delta_0}=0.6082,
\end{equation}
\cite{tanaka} which is the threshold value of the LOFF window 
along the transition line, 
where $\Delta_0$ is the gap at $T=0$ in the absence of the mismatch and 
$\frac{\Delta_0}{k_BT_0}=\pi e^{-\gamma_E}$, following the standard 
BCS relation.

What we have considered so far is the response of the free energy of a 2SC 
or a g2SC to a virtual displacement
of a small net momentum $\vec q$ in the direction of the simple LOFF pairing. 
The condition of the corresponding instability is the same as 
the chromomagnetic instability associated with the rotated 8-th 
gluon. The true minimum of 
the free energy with the LOFF pairing requires that
\begin{equation}
\vec\nabla_{\vec q}{\cal F}=0
\label{loff_min}
\end{equation}
and that $\frac{\partial^2{\cal F}}{\partial q_i\partial q_j}$ 
is non negative. 
The existence of the solution has been demonstrated 
in \cite{tanaka} for an electronic 
system ( though the gradient function (\ref{loff_min}) at $T=0$ 
may be discontinuous at the 
value of $q$ when the branch of the gapped excitations 
at $q=0$ become gapless). It
was found that the value of $q$ of the solution away from 
the transition temperature 
becomes quickly comparable to the inverse coherence
length at zero
temperature. An important property of the minimum is that the electric current 
vanishes there and consequently
there is no induced magnetic field. For the color
superconductor considered here, the condition (\ref{loff_min}) implies 
the vanishing tadpole diagram, i.e. 
\begin{equation}
<\bar\psi\vec\gamma\psi>=0.
\end{equation}
What about color and electric currents? In order to answer this 
question, we write down the color-electric current in terms of the exact 
quark propagator at finite $\vec q$ and $\delta$, i.e.
\begin{equation}
\vec J_l=-k_B T\sum_{\nu}\int\frac{d^3\vec p}{2\pi^3}
Tr\vec\Gamma^l{\cal S}_{\vec q}(P)
\end{equation}
where
\begin{equation}
{\cal S}_{\vec q}^{-1}(P)=\left( \begin{array}{cc} 
/\kern-7pt P-i\vec\gamma\cdot\vec q+\bar\mu\gamma_4
-\delta\gamma_4\tau_3+\delta^\prime\lambda_8 & \Delta\gamma_5\lambda_2 \\ 
-\Delta\gamma_5\lambda_2 & /\kern-7pt P+i\vec\gamma\cdot\vec q
-\bar\mu\gamma_4-\delta\gamma_4\tau_3-\delta^\prime\lambda_8
\end{array} \right).
\label{prop_gap_loff}
\end{equation}
Again, the isospin part of the charge operator contributes only a total 
derivative because of the relation
\begin{equation}
Tr\gamma_j\tau_3{\cal S}_{\vec q}(P)=
=-i\frac{\partial}{\partial p}_j\ln{\rm det}
{\cal S}_{\vec q}(P)\vert_{\tau_3=1}
+i\frac{\partial}{\partial p}_j\ln{\rm det}
{\cal S}_{\vec q}(P)\vert_{\tau_3=-1}
\end{equation}
and may be omitted. The residual $SU(2)\times{\cal U}(1)$ symmetry implies that
$\vec J_l=0$ ($l\neq 8$) for all $q$ and
\begin{equation}
\vec J_8=\frac{1}{6}\sqrt{3g^2+e^2}<\bar\psi\vec\gamma\psi>=0.
\end{equation}
for the value of $q$ that satisfies eq. (\ref{loff_min}. 
Therefore all currents vanish at the LOFF
minimum. While the chromomagnetic instability 
associated with the 8th gluon has been
removed, we need to check if the 
chromomagnetic instability associated with the other color components 
vanishes as well.

In principle, there exist several directions that
lead to a LOFF state corresponding to
imaginary Meissner masses 
for different gluons, along which a virtual displacement lowers the free 
energy of the condensate. This amounts in replacing the
transformation (\ref{loff}) by
\begin{equation} 
\chi(\vec r)=e^{i{\cal T}\vec q\cdot\vec r}\psi(\vec r)
\label{loff_1}
\end{equation}
where ${\cal T}$ is the combination of the group generators
that corresponds to the
negative eigenvalue of the mass square matrix $(m^2)^{l^\prime l}$. 
This is true both for
the other gluons ( $l=4,5,6,7$ ) of a 2SC superconductor
and for the chromomagnetic 
instability of a gCFL superconductor. It is not clear, however, 
if the condition (\ref{loff_min})
guarantees the vanishing of all the components of
the color current. If not, the true minimum
will be considerably more complicated than
the LOFF state since a nonzero expectation 
value of the gluon field will be
induced.

Before concluding this paper, we would like to argue that the charge 
neutrality condition may be 
implemented in the LOFF state as long as it  can be implemented 
in the g2SC superconductor. The gap equation and the charge 
neutrality condition can be written as
\begin{equation}
\frac{\partial{\cal F}}{\partial\Delta}=\frac{\partial{\cal F}}
{\partial\delta}=0
\label{cond}
\end{equation}
If we treat the second term of (\ref{pert}) perturbatively,
the deformation of the 
solution $(\Delta$, $\delta)$ to (\ref{cond}) from the one
at $q=0$ is of order $O(q^2)$ 
provided that the determinant
\begin{equation}
\left( \begin{array}{cc}
\frac{\partial^2{\cal F}_0}{\partial\Delta^2}&
\frac{\partial^2{\cal F}_0}{\partial\Delta\partial\delta} \\
\frac{\partial^2{\cal F}_0}{\partial\Delta\partial\delta} &
\frac{\partial^2{\cal F}_0}{\partial\delta^2}
\end{array} \right) \neq 0
\label{jacobian}
\end{equation}
at $q=0$. The numerical solution of \cite{shovkovy} indicates
that the stationary point defined
by (\ref{cond}) at $q=0$ and $T=0$ is a saddle point with respect
to $\Delta$ and $\delta$,
in which case the condition (\ref{jacobian}) is true. Therefore
the impact of the
charge neutrality condition
on the free energy for a small nonzero $q$ is of order $O(q^4)$
and as a result
cannot compete with the $q^2$ term
of (\ref{pert}). As $q$ gradually increases from zero, eqs.(\ref{cond})
defines a charge neutral trajectory in the three dimensional 
parameter space of
$\Delta$, $\delta$ and $q$ that intersects with
the plane of $q=0$ at the solution of
Ref. \cite{shovkovy}. Since the free energy is a contunuous
function and is bounded from below,
there exists a minimum at $q\neq 0$ along the
trajectory if $\kappa$ or equivalently $m_8^2$
is negative.

In this letter, we have explored the relation between the newly discovered 
chromomagnetic instability of a homogeneous color superconductor and the 
tendency towards a LOFF pairing by providing a net momentum to a Cooper pair 
to the free energy landscape. Our procedure is similar to that of Wu and Yip 
for a nonrelativistic superfluid \cite{wu}.
We showed that whenever the square of the Meissner mass of
the 8th gluon becomes negative there is a LOFF state with
lower free energy. Further investigation is still required
in order to decide
whether this LOFF state represents a true equilibrium phase of the quark 
matter at moderate baryon density. Future work will concentrate on
calculating
the Meissner masses at the LOFF minimum, examining the effect of 
charge neutrality, and comparing
the free energy of the LOFF minimun to the free energy of
other exotic phases proposed in the literature. We hope to report
on these issues in the near future.  

\section*{Acknowledgments}
We would like to thank M. Haung and I. Shovkovy for discussions and to thank 
M. Alford for bringing the reference \cite{rebby} to our attention. 
This work is supported in 
part by the US Department of Energy under grants
DE-FG02-91ER40651-TASKB.



\begin{thebibliography}{99}
\bibitem{color}
For early works, see: D.\ Bailin and A.\ Love, Phys.\ Rept.\ 107 (1984) 325,
and references therein.
\bibitem{reviews} K.\ Rajagopal and F.\ Wilczek, 
in B.L.\ Ioffe Festschrift, {\it At the Frontier of Particle
Physics/Handbook of QCD}, M.\ Shifman ed., 2061 (World Scientific 2001);
M.\ Alford, Ann.\ Rev.\ Nucl.\ Part.\ Sci.\ 51 (2001) 131;
T.\ Sch\"afer, hep-ph/0304281; D. H. Rischke, Prog. Part. Nucl. Phys., 
{\bf 52}, 197 (2004);
H.-c.\ Ren, hep-ph/0404074, M. \ Huang, hep-ph/0409167;
I. \ Shovkovy, nucl-th/0410091 and references therein.
\bibitem{alford} M.\ G. Alford, K.\ Rajagopal and F.\ Wilczek,
Nucl. \ Phys. \ B537 (1999) 443.
\bibitem{loff} M.\ G.\ Alford, J.\ A.\ Bowers 
and K.\ Rajagopal, Phys.\ Rev.\ D63 (2001)
074016; A.\ K.\ Leibovich, K.\ Rajagopal and E.\ Shuster, Phys. \ Rev. \ D64 (2001) 094005; 
I.\ Giannakis, J.\ T.\ Liu and H.\ C.\ Ren,
Phys.\ Rev.\ D66 (2002)
031501; D. K. Hong, hep-ph0107017; 
R. \ Casalbuoni, G.\ Nardulli, Rev.\ Mod.\ Phys., 76, (2004) 263.
\bibitem{shovkovy} I.\ Shovkovy and M.\ Huang, 
Phys.\ Lett.\ B564 (2003) 205; M. Huang and I. Shovkovy, Nucl.\ Phys.\ A729
(2003) 835; A. Mishra and H. Mishra, Phys. Rev. D 69, (2004) 014014.
\bibitem{chao} M. \ Huang, P. \ F. \ Zhuang 
and W. \ Q. \ Chao, Phys. \  Rev.  D67
(2003) 065015; Jinfeng Liao and Pengfei
Zhuang, Phys.\ Rev. \ D68 (2003) 114016.
\bibitem{kouvaris}
M. \ Alford, C. \ Kouvaris and K. \ Rajagopal, Phys. \ Rev. \ Lett.
92 (2004) 222001.
\bibitem{bedaque} P.\ F.\ Bedaque, 
H. \ Caldas and G. \ Rupak, Phys. \ Rev. \ Lett. 91 (2003)
247002; H. \ Caldas, Phys. \ Rev. \ A 69 (2004) 063602.
\bibitem{ruester}
K. Iida, T. Matsuura, M. Tachibana and T. Hatsuda, Phys. \ Rev. \ Lett. 93 
(2004) 132001;
B. \ Ruester, I. \ Shovkovy and D. \ Rischke, Nucl. \ Phys. A, 743, (2003) 127.
\bibitem{sarma} G.\ Sarma, \ Phys. \ Chem. \  Solid. 24 (1963) 1029.
\bibitem{liu} W. V. \ Liu and F.\ Wilczek, \ Phys. \ Rev. \ Lett. 90 (2003)
047002; E. Gubankova, W. V. Liu and F. Wilczek, Phys. \ Rev. \ Lett. 91
(2003) 032001.
\bibitem{wu} S. \ Wu and S. \ Yip, \ Phys. \ Rev. A67 (2003) 053603.
\bibitem{huang} M.\ Huang and I.\ Shovkovy {\it '' Screening masses 
in neutral two 
flavor color superconductor ''} hep-ph/0408268; M.\ Huang and I.\ Shovkovy, 
Phys.\ Rev.\ D70 (2004) 051501(R).
\bibitem{casal} R.\ Casalbuoni, R.\ Gatto, M.\ Mannarelli, G.\ Nardulli
and M.\ Raggieri {\it '' Meissner masses in the gCFL phase of QCD ''},
hep-ph/0410401.
\bibitem{rebby} S. \ Rebby and G. \ Rupak, {\it '' Phase structure of 2-flavor 
quark matter: heterogeneous superconductor ''} nucl-th/0405054.
\bibitem{tanaka} S. \ Takada and T. \ Izuyama, Prog. \ Theor. \
Phys. 41 (1969) 635.
\bibitem{hzc}
M. \ Huang, P. \ F. \ Zhuang and W. \ Q. \ Chao, Phys. \ Rev. D 65 (2002)
076012. 
\bibitem{mixing} M.\ Alford, J.\ Berges and K.\ Rajagopal, Nucl.\ Phys. 
\ B571 (2000) 269; E. V. Gorbar, Phys. Rev. D 62 (2000) 014007.
\bibitem{iida} K. \ Iida and G. \ Baym, Phys. \ Rev. \ D 63 (2001)
074018; K. \ Iida and G. \ Baym, Phys. \ Rev. D 66 (2002) 014015.
\bibitem{GR} I.\ Giannakis and H.-C.\ Ren, Phys. \ Rev. \ D 65 (2002)
054017; I.\ Giannakis and H.-C.\ Ren, Nucl. \ Phys. B669, (2003) 462. 
\end{thebibliography}
\end{document}